# The case for a non-expanding universe

Antonio Alfonso-Faus

E.U.I.T. Aeronáutica, Plaza Cardenal Cisneros s/n, 28040 Madrid, SPAIN

E-mail: aalfonsofaus@yahoo.es

**Abstract. –** We present the results of two empirical constancies: the fine structure constant ($\alpha$) and the Rydberg constant ($R_y$). When the speed of light $c$ is taken away from $\alpha$, as shown elsewhere, the constancy of $\alpha$ implies the constancy of the ratio $e^2/\hbar$, $e$ the charge of the electron and $\hbar$ Planck´s constant. This forces the charge of the electron $e$ to be constant as long as the action $\hbar$ (an angular momentum) is a true constant too. Then the constancy of $R_y$, the Rydberg expression, implies that the momentum $mc$ is also a true constant. This is just the second law of Newton. The Compton wavelength, $\hbar/mc$, is then a true constant and there is no expansion at the quantum mechanical level. General relativity then predicts that the universe is not expanding. It is the only solution for cosmology. The time variation of the speed of light explains the observed red shift.

## 1. - Introduction.

Within a time scale of the order of the age of the universe the fine structure constant α appears to be a true universal constant, Chand et al. [1]. The reported observations in the past of a time variation for this constant are meaningless at this scale, Web et al. [2]. The same happens for the Rydberg constant $R_y$, Peik et al. [3]. Given the constancy of α the possibility of time variations for the speed of light c has driven speculations on the possible time variation of the Planck´s constant $\hbar$, and/or the electronic charge $e$. However it has been shown, Alfonso-Faus [4], that the relativity principle implies that α does not contain the speed of light explicitly in this physical context. Then the assumed and observed (radioactive decay times depend on the $7^{th}$ power of $\hbar$, and no time variation has been observed) constancy of the Planck´s constant $\hbar$ forces to assume also that the electronic charge $e$ is a true constant.

The constancy of the $R_y$ together with the constancy of α, implies the constancy of the momentum *mc, m* the mass of the electron. This momentum *mc* is proportional to any momentum *mv*, because the ratio *v/c* must be constant with cosmological time. Otherwise the relativity formulae would be time dependent, which is not observed. Then both constancies, $R_y$ and α, imply the conservation of momentum in the universe. This is Newton´s second law. But under this new view we generalize the condition for the speed of light: *Any possible cosmological time variation of the speed of light necessarily implies a cosmological time variation of the mass m,* Alfonso-Faus, [5]. According to general relativity the momentum is inversely proportional to the cosmological scale factor R, Harrison [6]. Then the constancy of momentum necessarily implies that the cosmological scale factor is constant. The Universe is not expanding. It is the time variation of the speed of light that causes the observed red shift. There is no theoretical way-out for this conclusion. If one believes in the laboratory measurements and the validity of the general relativity theory the universe is not expanding and the speed of light is decreasing with time. Of course the masses are increasing with cosmological time (a Mass-Boom).

## 2. - The fine structure constant α.

The fine structure constant is given by the expression

$$\alpha = \frac{e^2}{4\pi\varepsilon\hbar c} \tag{1}$$

Where *e* is the electron charge, *ε* the vacuum permittivity, *ħ* the Planck´s constant and c the speed of light. It has been shown that *α* depends very little upon the red shift, Webb et al. [2]. It may even be an absolute constant, Chad et al. [1]. Assuming that the ratio $e^2/\hbar$ in (1) is constant then any possible small time variation in α would be associated with a time variation in the product *εc*. However we have shown, following the principle of relativity, Alfonso-Faus [4], that the value of the vacuum permittivity is

$$\varepsilon = \frac{1}{c} \tag{2}$$

And therefore the speed of light is not explicit in the formulation of α as given in (1). The new formulation is, Alfonso-Faus [4]

$$\alpha = \frac{e^2}{4\pi\hbar} \tag{3}$$

The conclusion is that the constancy of α is the constancy of the ratio $e^2/\hbar$. The constancy of $\hbar$, the action and quantum of angular momentum, is well known from the constancy of the radioactive decay times. The theoretical formulation of these times gives a power of 7 dependence on $\hbar$. Any small time variation in $\hbar$ would certainly have been observed, and this is not the case. Then the electron charge $e$ is a universal constant too. We conclude the following

*The observed constancy of α gives the constancy of the ratio* $e^2/\hbar$

*Then the observed constancy of* $\hbar$ *gives the constancy of e*

According to observations we can reasonable be sure that the three parameters $\alpha$, $\hbar$ and $e$ are constants of nature to a very high accuracy. This means that they are practically absolute universal constants within the time scale of the age of the universe.

**3. - The Rydberg constant $R_y$.**

The Rydberg constant is known to be a true absolute constant with a very high accuracy, Peik et al. [3]. It s given by

$$R_y = \frac{mc}{4\pi\hbar}\alpha^2 \tag{4}$$

where $m$ is the mass of the electron. The constancy of $R_y$ , $\alpha$ and $\hbar$ gives the constancy of the product

$$mc = constant \tag{5}$$

This is in fact the constancy of the total average momentum content mc, in this case for the electron. In order to preserve the principle of relativity with all its formulation, that depends on the dimensionless ratio of speeds $v/c$, there cannot be any time variation in this ratio and this is what is observed. Then the relation (5) is immediately converted to

$$mv = constant \tag{6}$$

In the absence of any interaction the momentum is the most known conserved physical property in the universe. And, of course, it is so predicted by the second law of Newton.

*The observed constancy of $R_y$ gives the constancy of momentum*

**4. - The Compton wavelength $\lambda_c$.**

It is well known that the Compton wavelength $\lambda_c$ of a mass *m* is given by

$$\lambda_c = \frac{\hbar}{mc} \tag{7}$$

and that this length gives an order of magnitude of the size of the object of mass $m$. Since theoretical arguments, as well as observations, give constancy for $\hbar$ with cosmological time, from (5) we arrive at the conclusion that the Compton wavelength is also a true constant. The quantum world is not expanding. And the same happens with the de Broglie wavelength $\hbar/mv$.

Harrison [6] has presented a very interesting thought experiment, the so called cosmic-box. The important point here is that from this experiment as well as general relativity, Harrison [6], the momentum of a quantum particle *mv* is inversely proportional to the cosmological scale factor R:

$$mv \sim \frac{1}{R} \tag{8}$$

There is no way-out of the conclusion that the constancy of momentum forces the constancy of the cosmological scale factor R, and therefore *the universe is not expanding.*

Since R is constant, the wavelengths $\lambda$ of photons are constant too. The observed red shift from distant galaxies is proportional to the measured frequency of the photons $\nu = c/\lambda$. The usual argument runs as follows: given the observed dependence of $\nu$ with distance, and therefore with cosmological time, a constant *c* implies an increasing $\lambda$ with cosmological time. This corresponds directly to the interpretation that the universe is expanding, that the scale factor R, proportional to $\lambda$, is increasing with time. In our case, having shown that $\lambda$ is constant, the only way-out is to ascribe to the speed of light *c* a decrease with time. Close to the origin of the universe this speed must have been enormous and this makes the well known horizon paradox disappear. A new cosmological model appears to be necessary as a new frame of work based on these findings.

The decrease of *c with time* has been analyzed elsewhere, Alfonso-Faus [5]

with the result

$$c = HL \tag{9}$$

Here $H$ is the Hubble constant $H \approx 1/t$ and $L$ the constant size of the universe. Then we have for the speed of light $c$ the result

$$c \approx \frac{L}{t} \tag{10}$$

*The constancy of momentum implies that the universe is not expanding*

*The observed red shift implies that the speed of light decreases with cosmological time, and therefore that the masses increase with time*

**5. – Conclusion.**

Theory and experiment combine to predict, with high accuracy, the constancies of the fine structure constant $\alpha$, the Planck´s constant $\hbar$, the charge of the electron $e$, the linear momentum $mv$ and the scale factor of the universe R. The red shift from distant galaxies is better interpreted as a result of the decrease of the speed of light. An increase of mass with cosmological time is also the immediate consequence. The stability of a non-expanding universe may be achieved by the equilibrium between expanding electrical forces and contracting gravitational ones, Alfonso-Faus [7].

**8. – References.**